\documentclass[11pt,a4paper, showpreprintnumber]{article}
\usepackage{jheppub_r}

\newcommand{\ba}{\begin{eqnarray}}
\newcommand{\ea}{\end{eqnarray}}
\newcommand{\Nrel}{N_\mathrm{rel}}
\newcommand{\Ndec}{N_\mathrm{dec}}

\newcommand{\fb}[2]{\left(\frac{#1}{#2}\right)}
\newcommand{\sqb}[2]{\sqrt{\frac{#1}{#2}}}
\def\GeV{{\rm \ GeV}}

\title{Spectator field dynamics in de Sitter and curvaton initial conditions}

\author[a,b]{Kari Enqvist,}
\author[a,b]{Rose N. Lerner,}
\author[c]{Olli Taanila}
\author[d]{and Anders Tranberg}
\affiliation[a]{University of Helsinki, P.O. Box 64, FI-00014, Helsinki, Finland.}
\affiliation[b]{Helsinki Institute of Physics, P.O. Box 64, FI-00014, Helsinki, Finland.}
\affiliation[c]{Faculty of Physics, University of Bielefeld, D-33615 Bielefeld, Germany}
\affiliation[d]{Niels Bohr International Academy, Niels Bohr Institute and Discovery Center, Blegdamsvej 17, DK-2100, Copenhagen, Denmark}

\emailAdd{kari.enqvist@helsinki.fi}
\emailAdd{rose.lerner@helsinki.fi}
\emailAdd{olli.taanila@iki.fi}
\emailAdd{antranbe@nbi.dk}

\preprint{BI-TP-2012/17, HIP-2012-12/TH}

\abstract{We investigate the stochastic behaviour of long wavelength modes of light spectator scalar fields during inflation. When starting from a classical field value, the probability distribution for the spectator both spreads out and moves towards an equilibrium distribution. We study the timescales for a mixed quadratic and quartic potential. The timescale of equilibration depends on the parameters of the model, and can be surprisingly large, even much more than thousands of $e$-folds. These results imply that the initial conditions for spectator fields are not automatically erased during inflation. Applying the results to the curvaton model, we calculate the probability distribution of the curvature perturbation and discuss `typical' Universes.}

\begin{document}
\maketitle

%%%%%%%%%%%%%%%%%%%%%%%%%%%%%%%%%%%%%%%%%
\section{Introduction}
\label{sec:intro}
%%%%%%%%%%%%%%%%%%%%%%%%%%%%%%%%%%%%%%%%%

Scalar fields can play a significant role in the evolution of the very early universe. The inflaton itself is an obvious example, since it drives the large-scale dynamics of the entire Universe. Scalar fields feature in many models motivated by particle physics, and are often called moduli fields. Even if the scalar fields do not drive inflation, they can still play a significant role in various physical processes. We call this type of field spectators, and denote them as $\sigma$. The curvaton \cite{curvorig} is a particular example of a spectator field, which after inflation gives rise to the observed curvature perturbation $\zeta$ and has been discussed extensively in the literature \cite{morecurvaton}. Other examples are provided by the quintessence field \cite{Caldwell:1997ii}, and by the MSSM flat direction fields, which can be considered as light spectators during inflation with interesting fluctuation properties \cite{EFG}. There are also models of inflation, such as assisted inflation \cite{Liddle:1998jc} and N-flation \cite{Dimopoulos:2005ac} in which multiple scalar fields collectively contribute to the inflationary expansion.

The interesting question is: what happens to the spectators during inflation, and what are their possible values at the end of inflation? We focus on light spectators, which are subject to inflationary fluctuations. The initial field values of the spectators may either have been determined by processes preceding inflation or by a phase transition during inflation (see for example \cite{subir}).  In all cases, during inflation the long-wavelength modes of the light spectators will be subject to a stochastic evolution that can be described by a Langevin equation and the ensuing Fokker-Planck equation, which yields the time evolution of the
spectator probability distribution \cite{Starobinsky86} (see also e.g.~\cite{Nakao:1988yi, Stewart91} for some early references on the subject). This is the approach we adopt in order to study the dynamics of spectator fields during inflation.

Formal solutions for the behaviour of a light, self-interacting field in a de Sitter background have been given by Starobinsky and Yokoyama \cite{Starobinsky}. There are also analytical expression for the stationary equilibrium solutions, which describe the equilibrium achieved after a sufficiently long period of inflation \cite{Starobinsky}. In the equilibrium limit, all information about the initial field configuration has been wiped out. To describe the capability of inflation to wipe out information from initial conditions, it is necessary to investigate what period is long enough to attain a distribution close to the equilibrium one. Thus, we assume that the spectator has a classical initial value $\sigma_0$ at the beginning of the final period of inflation; whether this value is attained before inflation or during inflation is irrelevant for our purposes. We then solve the time evolution of the distribution, and ask how many $e$-folds of inflation it takes to get close to the equilibrium distribution.

We concentrate on the spectator field potential $V=m^2\phi^2/2+\lambda\phi^4/4$. In the limit $\lambda \to 0$, the distribution is Gaussian and analytically solvable. However, the general case requires a numerical solution of the Fokker-Planck equation, which we show in section \ref{sec:FPsolution}. We discuss the evolution of the probability distributions as a function of the parameters of the potential and the number of $e$-folds $N$. We determine both the relaxation time, which measures the rate by which equilibrium is approached, and the decoherence time, which measures the spreading of the initial delta-peak like distribution\footnote{This is not decoherence in the quantum sense.}. We also discuss the evolution of the lowest $n$-point correlators and pay particular attention to the transient evolution that takes place before the system has time to decohere. An important conclusion is that the equilibrium behaviour may only be approached after a large number of $e$-folds, because the transient evolution can easily take hundreds or hundreds of thousands of $e$-folds, depending on the parameter values of the spectator potential.

After discussing the general evolution of spectator fields during inflation, we then focus on how this translates into a probability distribution of the curvature perturbation $\zeta$ in the cases where the spectator field is a curvaton (section \ref{sec:quarquad2}). We conclude in section \ref{sec:conc}.

%%%%%%%%%%%%%%%%%%%%%%%%%%%%%%%%%%%%%%%%%
\section{Behaviour of spectator fields}
\label{sec:FPsolution}
%%%%%%%%%%%%%%%%%%%%%%%%%%%%%%%%%%%%%%%%%

%%%%%%%%%%%%%%%%%%%%%%%%%%%%%%%%%%%%%%%
\subsection{The Fokker-Planck equation}
\label{sec:FPsetup}
%%%%%%%%%%%%%%%%%%%%%%%%%%%%%%%%%%%%%%%
Let us consider a light ($m \ll H_*$) real scalar spectator field $\sigma$ evolving in a de Sitter Universe. Inflation is sustained by some other physics, not necessarily a scalar inflaton. For simplicity, we assume that the Hubble rate remains constant throughout inflation. After integrating out the short-wavelength modes $k\ll H_*$ of the spectator field (using an appropriate window function), the evolution of the long-wavelength modes $\sigma$  can be approximated by a Langevin equation of the form
\ba
\dot{\sigma}=\frac{V'(\sigma)}{3H_*}+\xi(t),
\ea
where the random Gaussian noise $\xi(t)$ has the correlator
\ba
\langle \xi(t)\xi(t')\rangle=\delta(t-t')\frac{H_*^3}{8\pi^2}.
\ea
As a consequence, the evolution of the probability distribution of the spectator $\sigma$ can be shown to obey a Fokker-Planck equation, which reads \cite{Starobinsky}
\ba
\label{FPeq}
\dot P(\sigma,N)= \frac{1}{3H_*^2}V''(\sigma)P(\sigma,N)+\frac{1}{3H_*^2}V'(\sigma)P'(\sigma,N)+\frac{H_*^2}{8\pi^2}P''(\sigma,N),
\ea
where the dot is a derivative w.r.t the $e$-folds $N$, and the prime is a derivative w.r.t $\sigma$. Given the initial probability distribution, solving (\ref{FPeq}) yields the distribution for all $N$. In what follows we will assume a fixed classical initial value $\sigma_0$, essentially a delta-peak distribution $\sim \delta(\sigma-\sigma_0)$, which for numerical purposes will be taken to be a narrow Gaussian.

Let us further assume a potential given by
\ba
V(\sigma)=\frac{1}{2}m^2\sigma^2+\frac{1}{4}\lambda\sigma^4,
\ea
where $\lambda<1$. In the $N=\infty$ limit, the probability distribution will reach a stationary form, given by
\ba\label{equdist}
P(\infty,\sigma)=\mathcal{N}\exp\left(-\frac{8\pi^2}{3H^4_*}V(\sigma)\right)=
\frac{\exp\left[-\frac{8\pi^2}{3H^4_*}\left(\frac{1}{2}m^2\sigma^2+\frac{1}{4}\lambda\sigma^4\right)\right]}{\sqrt{\frac{m^2}{2\lambda}}\exp\left(\frac{m^4\pi^2}{3\lambda}\right)K_{1/4}\left(\frac{m^4\pi^2}{3\lambda}\right)},
\ea
where $K_n(x)$ is a Bessel function and $\mathcal{N}$ is a normalization constant. The equilibrium distribution (\ref{equdist}) is independent of the initial condition. The question then is: how long does it take for the distribution to reach near equilibrium, and what is the rate by which the initial delta-function spreads out?

%%%%%%%%%%%%%%%%%%%%%%%%%%%%%%%%%%%%%%%
\subsection{Quadratic potential}
\label{sec:quad}
%%%%%%%%%%%%%%%%%%%%%%%%%%%%%%%%%%%%%%%

%%%%%%%%%%%%%%%%%%%%%%%%%%%%%%%%%%%%%%%%%%%%%%%%%%%%%%%%%%%%%%%%%%%%%%%%
\begin{figure}
\begin{center}
\includegraphics[width=0.5\textwidth]{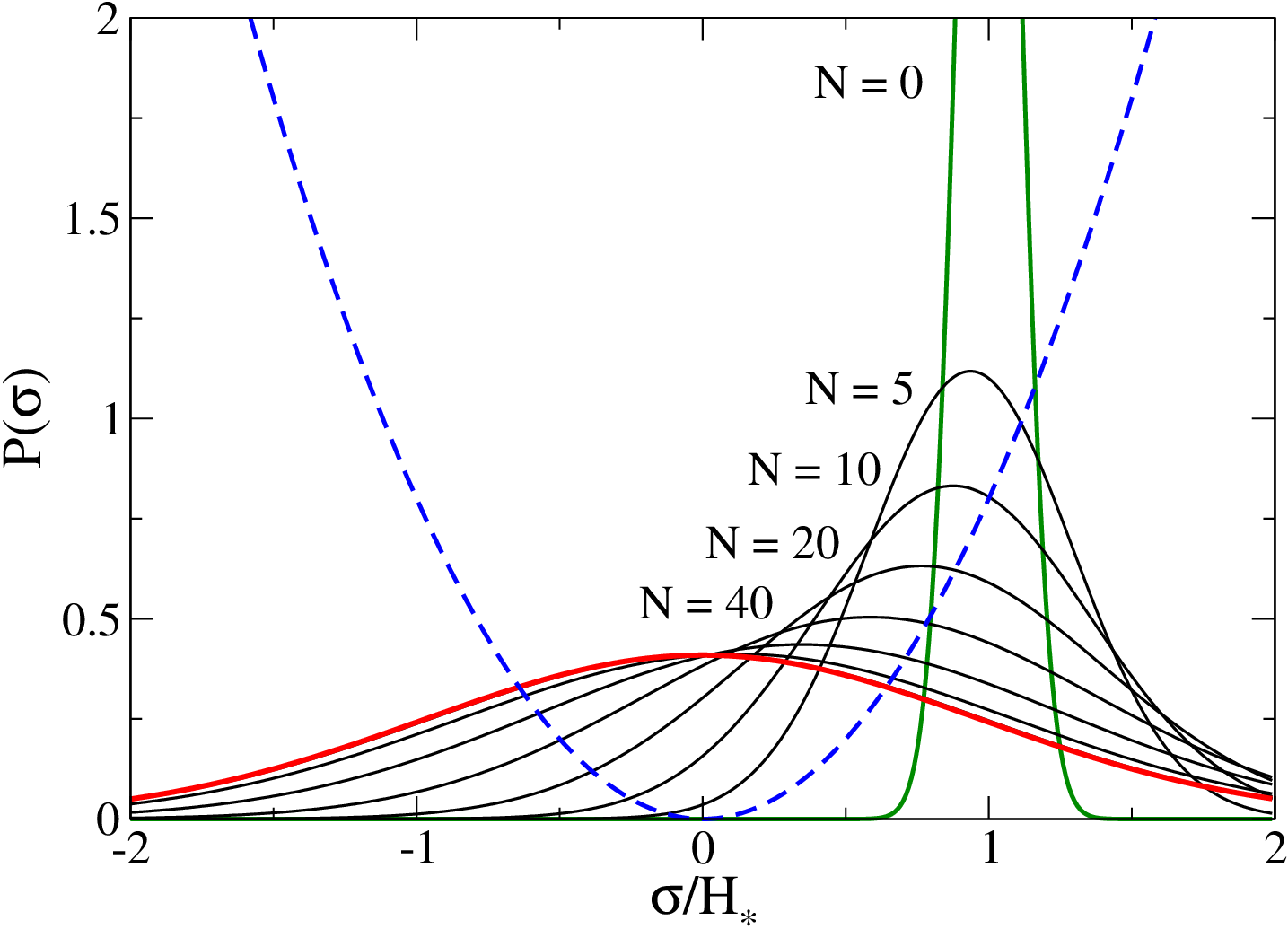}
\caption{Quadratic potential: evolution of the probability distribution starting from a Gaussian (green) converging to the equilibrium distribution (red). The blue dashed line is the potential. The parameters used are $\lambda=0$, $m/H_*=0.2$, $\sigma_0/H_*=1$, $w(0)/H_*=0.1$ which give $N_{rel} = 75 $ and $N_{dec} = 37.5$.}
\label{fig:quadevol}
\end{center}
\end{figure}
%%%%%%%%%%%%%%%%%%%%%%%%%%%%%%%%%%%%%%%%%%%%%%%%%%%%%%%%%%%%%%%%%%%%%%%%

Let us first consider the quadratic potential, i.e. the limit $\lambda=0$. This simplest case is solved analytically using the Gaussian ansatz
\ba
P(\sigma,N)=\frac{1}{\sqrt{2\pi w^2(N)}}\exp{\left(-\frac{(\sigma-\sigma_c(N))^2}{2H_*^2 w^2(N)}\right)}~.
\ea
Using (\ref{FPeq}) with arbitrarily narrow initial width, we determine the average $\sigma_c(N)$ and the width $w^2(N)$:
\ba
\label{eq:quad1}
\langle\sigma\rangle(N)&=&\sigma_c(N)=\sigma_c(0)\exp{\left(-\frac{m^2}{3H_*^2}N\right)},\\
\langle\sigma^2\rangle(N)-\sigma_c^2(N)&=&w^2(N)=\frac{3H_*^2}{8\pi^2m^2}-\left(\frac{3H_*^2}{8\pi^2m^2}-w^2(0)\right)\exp{\left(-\frac{2m^2}{3H_*^2}N\right)},
\label{eq:quad2}
\ea
where $\sigma_c(0)\equiv \sigma_0$ is the initial central value of the distribution.

From the exponents we read off two time-scales (or $N$-scales)
\ba
\label{eq:quadNs}
N_{\rm rel}=\frac{3H_*^2}{m^2},\qquad N_{\rm dec}=\frac{3H_*^2}{2m^2},
\ea
which we name the relaxation and decoherence time-scales, respectively. The relaxation time measures the rate of approach to the average field value at equilibrium, whereas the decoherence time measures the rate by which the initial narrow distribution broadens and spreads out towards the equilibrium width.

In Fig.~\ref{fig:quadevol} we show the evolution of the distribution starting from $\sigma_0/H_*=1$ and $w(0)/H_*=0.1$, with $m/H_*=0.2$, so that $N_{\rm rel}=2N_{\rm dec}=75$. We see how the initial peaked distribution (green) remains a Gaussian throughout and approaches the asymptotic form (red). The blue dashed line is the potential in some arbitrary normalization.

%%%%%%%%%%%%%%%%%%%%%%%%%%%%%%%%%%%%%%%
\subsection{Quartic potential}
\label{sec:quar}
%%%%%%%%%%%%%%%%%%%%%%%%%%%%%%%%%%%%%%%%

%%%%%%%%%%%%%%%%%%%%%%%%%%%%%%%%%%%%%%%%%%%%%%%%%%%%%%%%%%%%%%%%%%%%%%%%
\begin{figure}
\begin{center}
\includegraphics[width=0.5\textwidth]{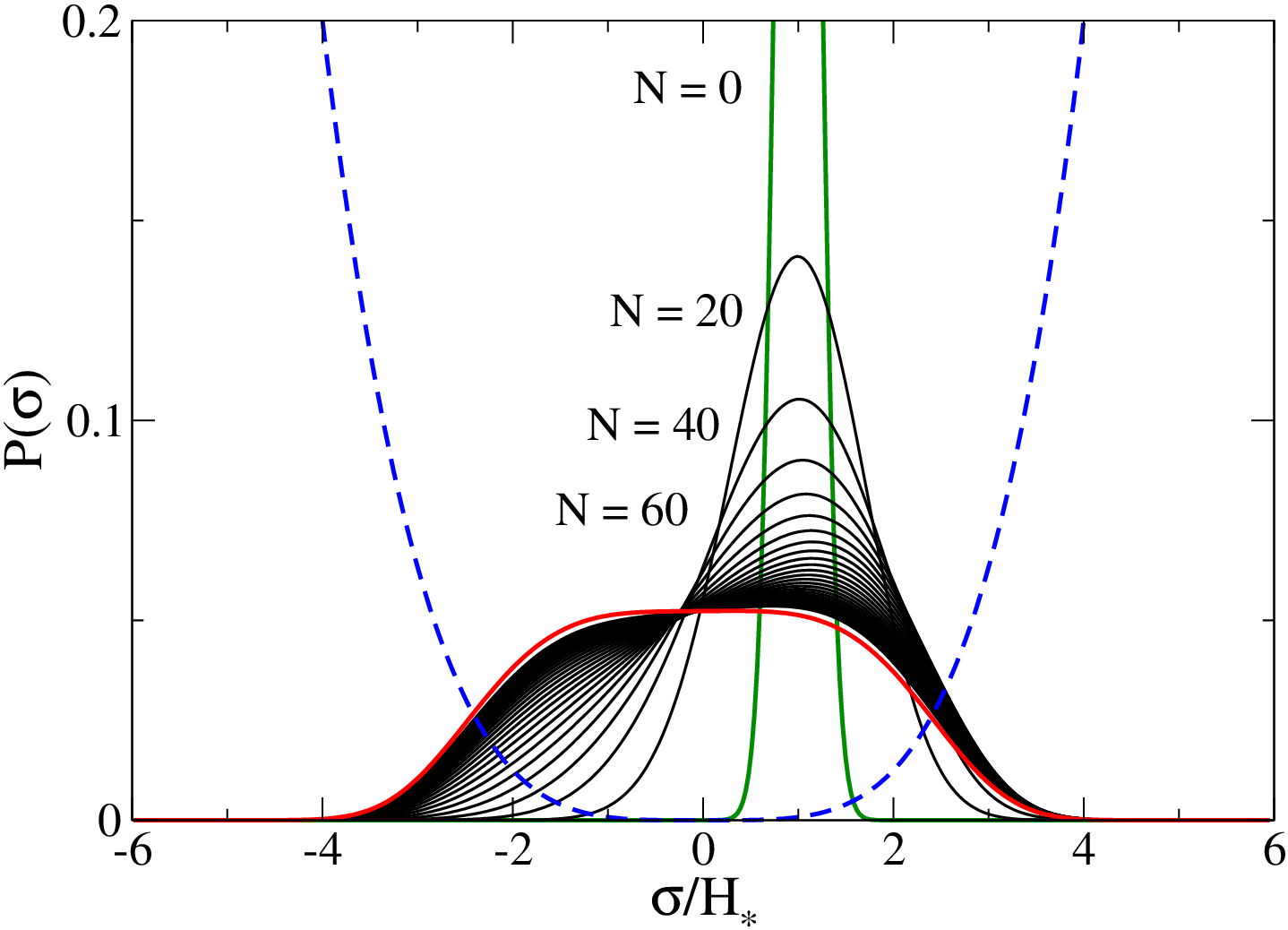}
\caption{Quartic potential: evolution of the probability distribution starting from a Gaussian (green) converging to the stationary distribution (red). The blue dashed line is the potential. The parameters used are $m=0$, $\lambda=0.003125$, $\sigma_0/H_*=1$ and $w(0)/H_*=0.1$ and the black lines are spaced by 20 $e$-folds.}
\label{fig:evol2}
\end{center}
\end{figure}
%%%%%%%%%%%%%%%%%%%%%%%%%%%%%%%%%%%%%%%%%%%%%%%%%%%%%%%%%%%%%%%%%%%%%%%%

%%%%%%%%%%%%%%%%%%%%%%%%%%%%%%%%%%%%%%%%%%%%%%%%%%%%%%%%%%%%%%%%%%%%%%%%
\begin{figure}
\begin{center}
\includegraphics[width=0.45\textwidth]{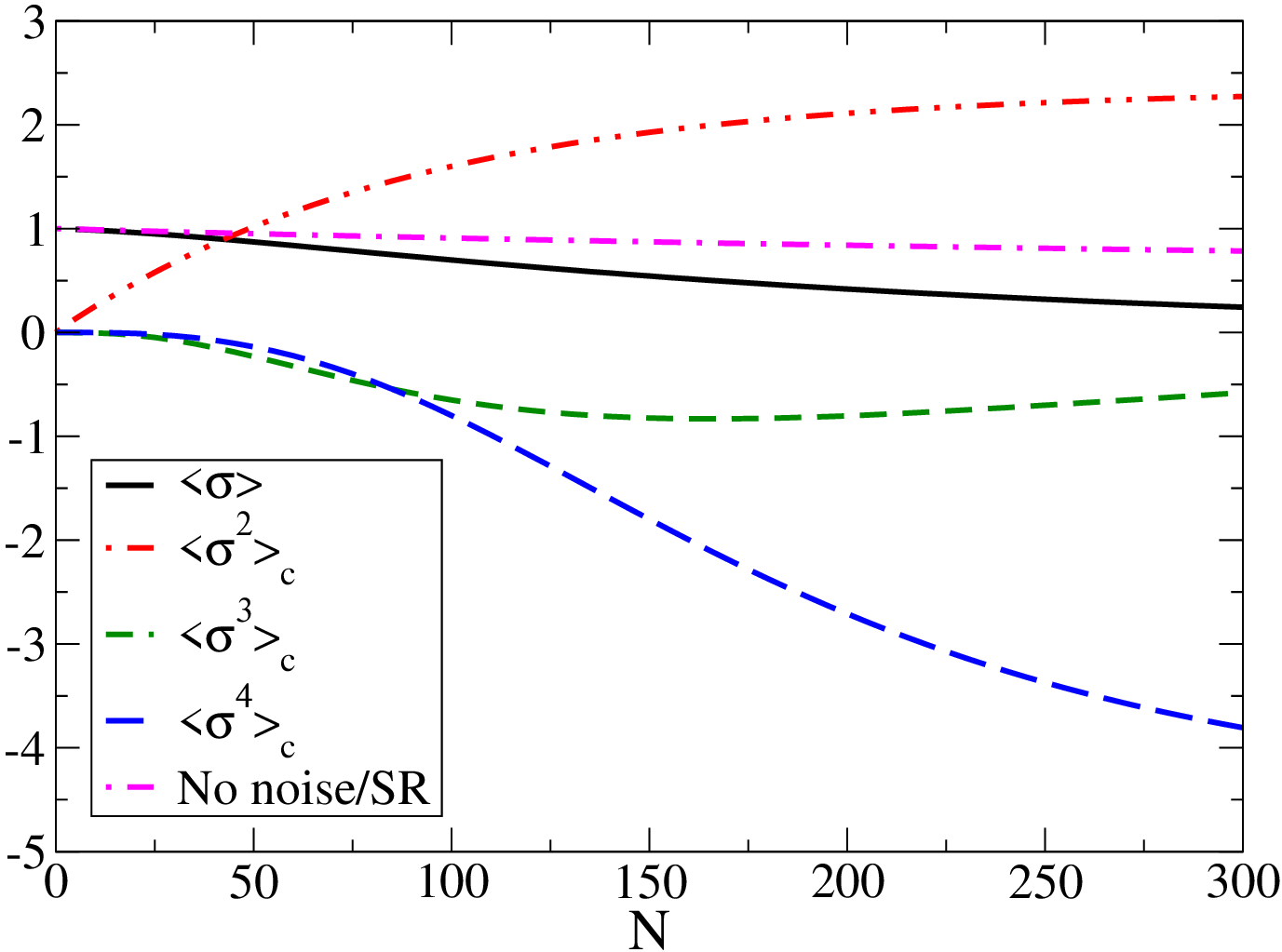}
\includegraphics[width=0.49\textwidth]{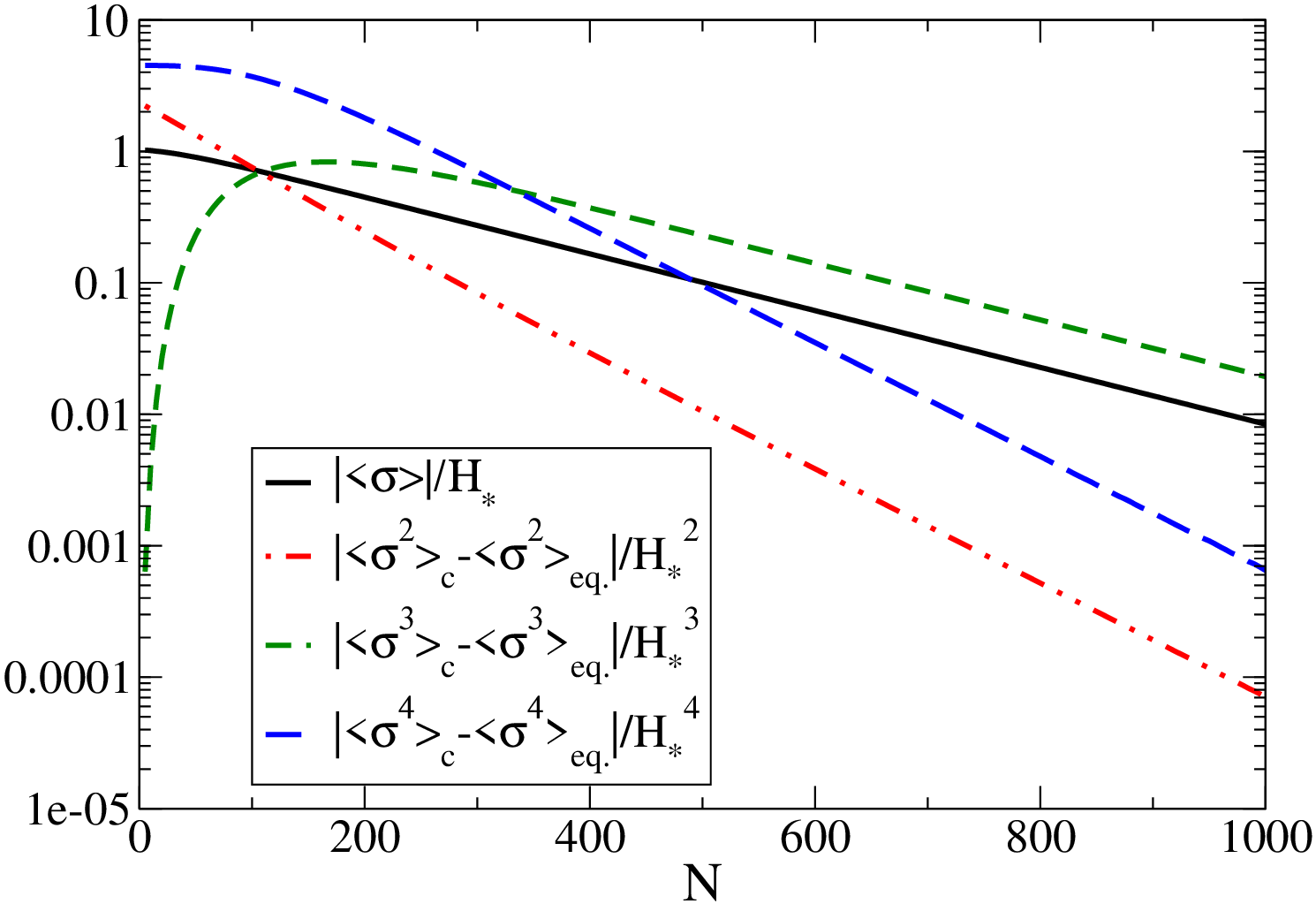}
\caption{Left: The connected 1, 2, 3 and 4-point correlators for $\lambda=0.03125$, $m=0$. Also shown is the noise-less, pure slow-roll evolution (magenta). Right: The deviation of the correlators from their equilibrium values. When the lines becomes straight, the correlators are exponentially approaching their equilibrium values. The transient behaviour at small $N$ depends on the initial conditions.}
\label{fig:allcorr0.1}
\end{center}
\end{figure}
%%%%%%%%%%%%%%%%%%%%%%%%%%%%%%%%%%%%%%%%%%%%%%%%%%%%%%%%%%%%%%%%%%%%%%%%

%%%%%%%%%%%%%%%%%%%%%%%%%%%%%%%%%%%%%%%%%%%%%%%%%%%%%%%%%%%%%%%%%%%%%%%
\begin{figure}
\begin{center}
\includegraphics[width=0.5\textwidth]{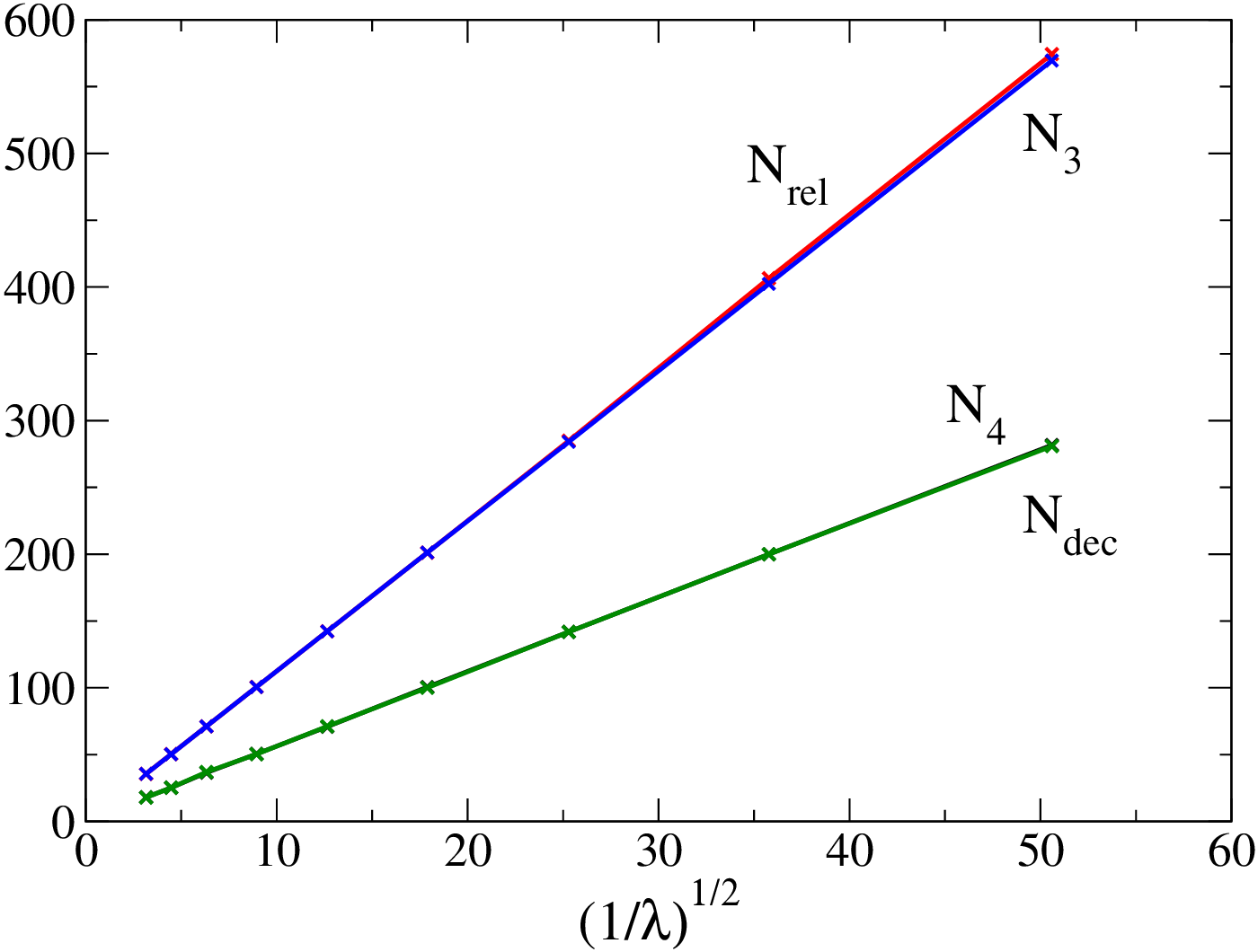}
\caption{Quartic potential: The dependence of the relaxation, decoherence and other timescales on $\lambda^{-1/2}$. Parameters used are $\sigma_0/H_*=1$, $m=0$ and $w(0)/H_*=0.1$. Note that the lines $N_{dec}$ and $N_4$ lie exactly on top of each other, as do the $N_3$ and $N_{rel}$ lines (within numerical errors).}
\label{fig:nreldec}
\end{center}
\end{figure}
%%%%%%%%%%%%%%%%%%%%%%%%%%%%%%%%%%%%%%%%%%%%%%%%%%%%%%%%%%%%%%%%%%%%%%%

The solution for the pure quartic potential ($m=0$) is more complicated compared to the solution for the pure quadratic potential. For the quadratic potential, the probability distribution is Gaussian at all times, with a mean and width that evolve in time. For the quartic case, this is no longer true. The initially narrow Gaussian distribution becomes distorted before settling down to its asymptotic form.

For the quartic potential we need to solve the Fokker-Planck equation numerically, and then extract the time-scales analogous to $\Nrel$ and $\Ndec$ in the quadratic case. For numerical stability reasons, the equation has to be solved using an implicit discretization. Also, a numerical solution limits the range of $\lambda$ and $m^2$ directly available to us; in particular, the small $\lambda$ region remains inaccessible numerically. Nevertheless, the results allow us to confidently extrapolate to small couplings.

We plot the evolution of the probability distribution in figure \ref{fig:evol2} at different values of $N$. The initial distribution is given by a narrow Gaussian with $w(0)/H_* = 0.1$, $\sigma_0/H_* = 1$ (green curve) where $m=0$ and $\lambda = 0.003125$. The grey lines are spaced by 20 $e$-folds. The initial Gaussian is transforming into the equilibrium distribution, which is not Gaussian in this case. Superficially, the evolution of the probability distribution function appears similar compared to the quadratic case of figure \ref{fig:quadevol}. However, there are interesting differences, as we now discuss.

To quantify the time dependence of this solution, we plot the connected 1, 2, 3 and 4-point correlators in figure \ref{fig:allcorr0.1}. The plot on the right hand side of figure \ref{fig:allcorr0.1} shows the difference between the correlators and their equilibrium values. The vertical axis is logarithmic and thus we can see that the $n$-point correlators are well approximated by exponential laws, since they appear linear in this logarithmic plot. To find the relaxation and decoherence times related to this exponential behaviour, we make a fit using the ansatz
\begin{equation}
C_n(N) = C_n(\infty)-(C_n(\infty)-C_n(0))e^{-\frac{N}{N_{n}}} \; .
\end{equation}
Here $C_n$ is the connected $n$-point correlator. Analogously to the quadratic case, we denote $N_1 = \Nrel$ and $N_2 = \Ndec$.

For different values of the coupling $\lambda$ we perform such exponential fits, also establishing the range of $N$  for which the exponential behaviour is realised. This gives us figure \ref{fig:nreldec}. We have chosen to rescale the abscissa to $\lambda^{-1/2}$, to illustrate the dependence of these timescales. Not only are they straight lines, but the odd order correlators have the same scaling. The even order correlators have another scaling, but one is twice the other, in complete analogy with the Gaussian case (equations \ref{eq:quad1}, \ref{eq:quad2}). The timescales are given by
\begin{eqnarray}
N_{\rm rel}=N_{3}\simeq \frac{11.3}{\sqrt{\lambda}} \; ,\\
N_{\rm dec}=N_{4}\simeq \frac{5.65}{\sqrt{\lambda}} \; .
\end{eqnarray}

Although the evolution of the $n$-point correlators is exponential for much of the evolution (i.e.~looks linear in figure \ref{fig:allcorr0.1}), during the first few hundred $e$-folds the evolution is non-exponential. This initial non-exponential behaviour we called \emph{transient} behaviour. Here the distribution develops a $3$-point correlator as it rolls down the potential. To a reasonable accuracy, we have found that the exponential behaviour begins around $N=N_{\rm dec}$ in the evolution. We have also verified that by the time the exponential regime has been reached, the time-scales $\Nrel$ and $\Ndec$ are independent of the starting distribution.

Although both the initial and the equilibrium distribution are symmetrical about the mean, the distribution can be quite skewed for the transition period, which can last hundreds, even tens of thousands of $e$-folds, depending on $\lambda$. The smaller $\lambda$, the longer the transition period. Therefore we conclude that even in the relatively simple quartic case, the light spectators are not generally well described by their equilibrium distribution. Rather, it would appear that for inflationary scenarios where inflation lasts only for $\mathcal{O}(100)$ $e$-folds, the distributions of light spectators are dominated by their value before inflation and subsequent transient behaviour.

%%%%%%%%%%%%%%%%%%%%%%%%%%%%%%%%%%%%%%%
\subsection{Mixed quartic and quadratic potential}
\label{sec:quarquad}
%%%%%%%%%%%%%%%%%%%%%%%%%%%%%%%%%%%%%%%%

%%%%%%%%%%%%%%%%%%%%%%%%%%%%%%%%%%%%%%%%%%%%%%%%%%%%%%%%%%%%%%%%%%%%%%%%
\begin{figure}
\begin{center}
\includegraphics[width=0.5\textwidth]{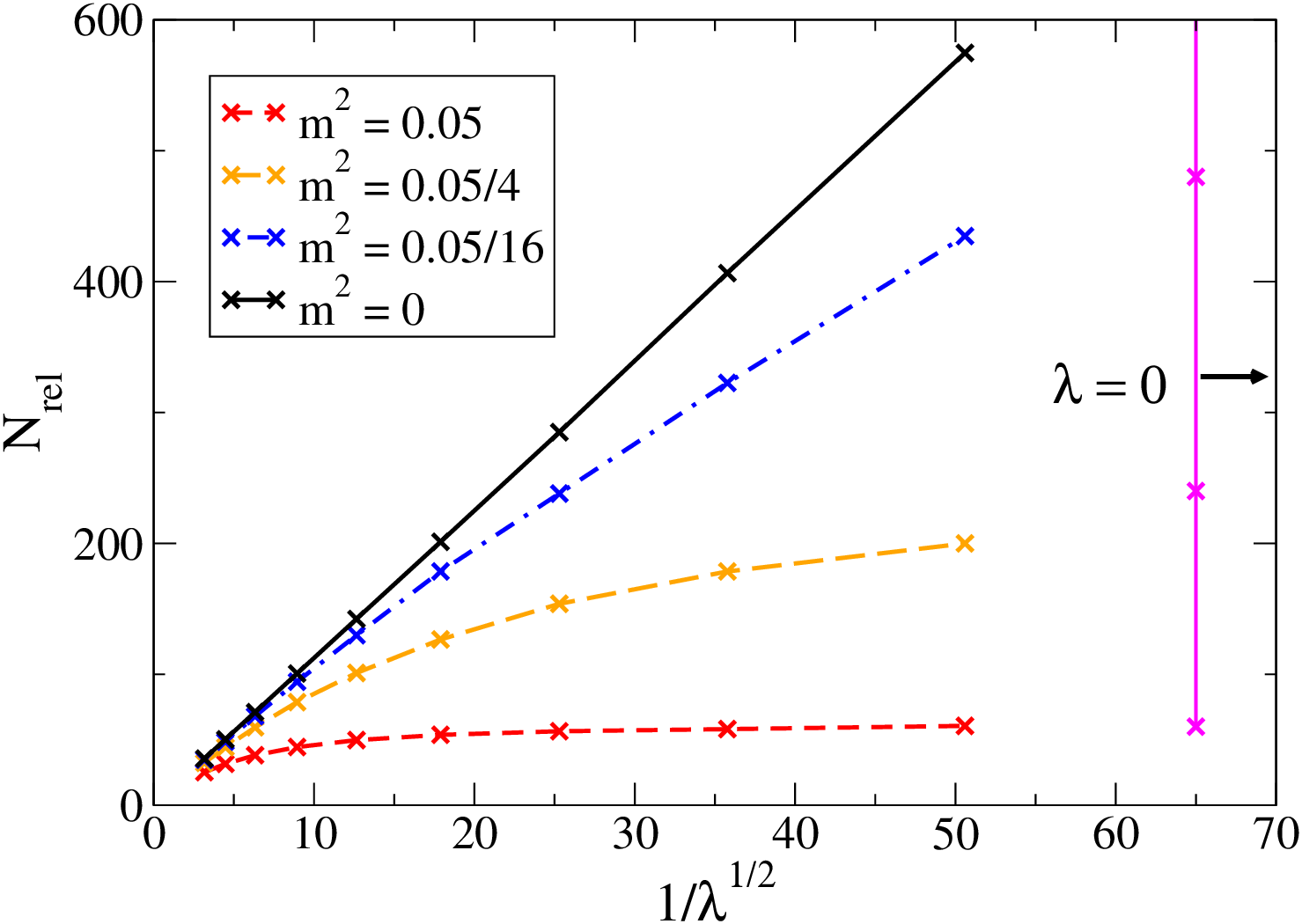}
\caption{Mixed quadratic and quartic potential: The dependence of the $N_{rel}$ on $\lambda^{-1/2}$ and $m^2$. Parameters used are $\sigma_0/H_*=1$ and $w(0)/H_*=0.1$. The upper solid line is $m^2 = 0$; $m^2$ increases in the lower lines. Note that these lines would asymptotically meet the points on the $\lambda=0$ line, which should be placed at $\infty$.}
\label{fig:Ns24}
\end{center}
\end{figure}
%%%%%%%%%%%%%%%%%%%%%%%%%%%%%%%%%%%%%%%%%%%%%%%%%%%%%%%%%%%%%%%%%%%%%%%%

For the full potential ($m^2>0$, $\lambda>0$) we again solve the Fokker-Planck equation and find an exponential regime for the behaviour of $\langle\sigma\rangle$ and $\langle\sigma^2\rangle_c$. Again we find that to a good approximation (i.e. to the numerical precision and the degree to which the evolution is strictly exponential), $N_{\rm rel}=2N_{\rm dec}$. In Fig.~\ref{fig:Ns24}, we again show $N_{\rm rel}$ with the abscissa $\lambda^{-1/2}$. We note that the dependence is no longer linear in $\lambda^{-1/2}$. The solid black line is the pure quartic $m^2=0$ result, which is a straight line. Increasing the mass to $m^2=0.05/16$, $0.05/4$ and $0.05$ gives the dashed lines. The solid (magenta) vertical line gives the pure quadratic result, and should be placed at $1/\lambda^{1/2} = \infty$. The dashed lines will asymptotically approach the points on this $\lambda = 0$ line, for small enough $\lambda$. When $\lambda$ decreases, the relaxation time becomes more and more dominated by the mass term. For smaller $m^2$ a similar behaviour can be found, where the relaxation time depends on $\lambda$ down to a certain value of the coupling, below which it approaches the $\lambda=0$ limit at that particular $m^2$. Thus, for either large $m^2$ {\em or} large $\lambda$, the relaxation time is short.

We emphasise that because of the noise term in the Fokker-Planck equation (the last term in \ref{FPeq}), the relaxation time is not simply a function of the ratio $m^2/\lambda$ as for the classical dynamics.

%%%%%%%%%%%%%%%%%%%%%%%%%%%%%%%%%%%%%%%
\section{Initial conditions for the curvaton}
\label{sec:quarquad2}
%%%%%%%%%%%%%%%%%%%%%%%%%%%%%%%%%%%%%%%%

%%%%%%%%%%%%%%%%%%%%%%%%%%%%%%%%%%%%%%%%%%%%%%
\subsection{Curvaton probability distribution}
\label{sec:zeta_quad}
%%%%%%%%%%%%%%%%%%%%%%%%%%%%%%%%%%%%%%%%%%%%%%%

During inflation all light spectators obtain an isocurvature perturbation; the curvaton is a specific type of spectator which survives inflation and whose isocurvature perturbation is eventually converted to the observed adiabatic curvature perturbation  $\zeta$. Once inflation has ended and the inflaton has decayed, the curvaton starts to oscillate in its potential. We assume that the Universe is radiation dominated at that point. The relative amplitude of the curvaton field perturbation is initially negligible, but is then enhanced so that at the time of curvaton decay, the perturbation can be imprinted on the decay products as the dominant adiabatic perturbation. At the time of the decay, the curvaton energy fraction may either dominate over the inflaton decay products or be subdominant; this has an effect on the magnitude of the generated curvature perturbation. Here we will denote the effective decay rate of the curvaton by $\Gamma$.

The simplest curvaton potential is a quadratic one. The perturbation amplitude and especially the non-Gaussianities of the perturbation may depend on the form of the potential, as discussed in \cite{selfinteract}. Here we do not aim at a detailed survey of the model parameters so that it suffices to consider the simplest quadratic case with $\lambda=0$. In that case, the curvature perturbation $\zeta$ is given by
\ba
\label{eq:zetafromsigma}
\zeta=r_{dec}\frac{H_*}{3\pi\sigma_*} = \left.\frac{3\rho_\sigma}{3\rho_\sigma+4\rho_{\rm rest}} \right|_{dec}\frac{H_*}{3\pi\sigma_*},
\ea
where $\sigma_*$ is the curvaton field value when the observable scales exit the horizon during inflation. The energy densities are evaluated at the time of curvaton decay and are given by
\ba
\rho_\sigma = 2.09 \fb{H_*}{m}^{3/2} \frac{m^2\sigma_{*}^2}{2} \fb{a_*}{a(t)}^3,\qquad \rho_{\rm rest}= 3M_{\rm Pl}^2H_*^2 \fb{a_*}{a(t)}^4,
\ea
because they have the equation of state of matter and radiation, respectively. The curvaton slow-rolls until $H(t)\simeq m$, which explains the factor $\propto \fb{H_*}{m}^{3/2}$; $\rho_{\rm rest}$ is assumed to be responsible the Hubble expansion $H_*$ prior to $\sigma_*$. We evaluate these quantities at curvaton decay, taken to be when  $1/\Gamma \simeq 1/H(t)$, where $\Gamma$ is the effective decay width and radiation domination is assumed.% $H(t) = 1/(2t)$.
This means that
\ba
\rho_\sigma|_{dec} \simeq m^{1/2}\,\sigma^2_*\,\Gamma^{3/2},\qquad
\rho_{\rm rest}|_{dec}=3 M_{\rm Pl}^2\Gamma^2
\ea
and thus we have
\ba
\label{eq:zetainv}
\zeta=\frac{1}{3\pi}\frac{\sigma_* H_*}{\sigma^2_*+4 M_{\rm Pl}^2\left(\frac{\Gamma}{m}\right)^{1/2}}.
\ea
Inverting this, we find
\ba
\label{sigpm}
\sigma_*^\pm=\frac{H_*}{6\pi\zeta}\left(1\pm\sqrt{1-\frac{144\pi^2 M_{\rm Pl}^2 \Gamma^{1/2}\zeta^2}{H_*^2 m^{1/2}}}\right).
\ea
For fixed curvature perturbation $\zeta$,  there are two solutions for $\sigma_*$, denoted here as ${\sigma}_*^+$  and ${\sigma}_*^-$, which correspond respectively to the dominant curvaton with $r_{\rm dec}\simeq 1$ and the subdominant curvaton with $r_{\rm dec}\ll 1$. Note also that there are no solutions for a particular $\zeta$ unless
\ba
\label{eq:critzeta}
\frac{144\pi^2 M_{\rm Pl}^2 \Gamma^{1/2}\zeta^2}{H_*^2 m^{1/2}} < 1.
\ea
As an example, we now set $\Gamma=10^{-15}\GeV$ and $H_*=10^{10}\GeV$ ($M_{\rm Pl}=2.435\times 10^{18} \GeV$ is the reduced Planck mass). For these parameters, we need $m>70$ TeV to get $\zeta \geq 10^{-5}$.

The probability distribution of $\zeta$ is
\ba
P(\zeta)&=&
P[\sigma_*^-(\zeta)]\left|\frac{d\sigma_*}{d\zeta}\right|_{\sigma_*^-(\zeta)}+
P[\sigma_*^+(\zeta)]\left|\frac{d\sigma_*}{d\zeta}\right|_{\sigma_*^+(\zeta)} \nonumber \\
& = & \sqb{4\pi m^2}{3H_*^2} \exp\fb{-m^2 \left(1-\sqrt{1-X\zeta^2}\right)^2}{27 \zeta^2 H_*^2} \frac{H_* \left(1-\sqrt{1-X\zeta^2}\right)}{6\pi \zeta^2 \sqrt{1-X\zeta^2}}  \nonumber \\
& & {} + \sqb{4\pi m^2}{3H_*^2} \exp\fb{-m^2 \left(1+\sqrt{1-X\zeta^2}\right)^2}{27 \zeta^2 H_*^2}\frac{H_* \left(1+\sqrt{1-X\zeta^2}\right)}{6\pi \zeta^2 \sqrt{1-X\zeta^2}},
\ea
where $X = \frac{144\pi^2 M_{\rm Pl}^2}{H_*^2}\sqb{\Gamma}{m}$, and the equations are valid only when \ref{eq:critzeta} is satisfied. The distribution has non-trivial behaviour, given by an interplay between the condition \ref{eq:critzeta}, the width of $P(\sigma)$ and the shape of $\zeta(\sigma)$.

%%%%%%%%%%%%%%%%%%%%%%%%%%%%%%%%%%%%%%%%%%%%%
\subsection{Typical Universes}
%%%%%%%%%%%%%%%%%%%%%%%%%%%%%%%%%%%%%%%%%%%

%
%%%%%%%%%%%%%%%%%%%%%%%%%%%%%%%%%%%%%%%%%%%%%%%%%%%%%%%%%%%%%%%%%%%%%%%%
\begin{figure}
\begin{center}
\includegraphics[width=0.5\textwidth]{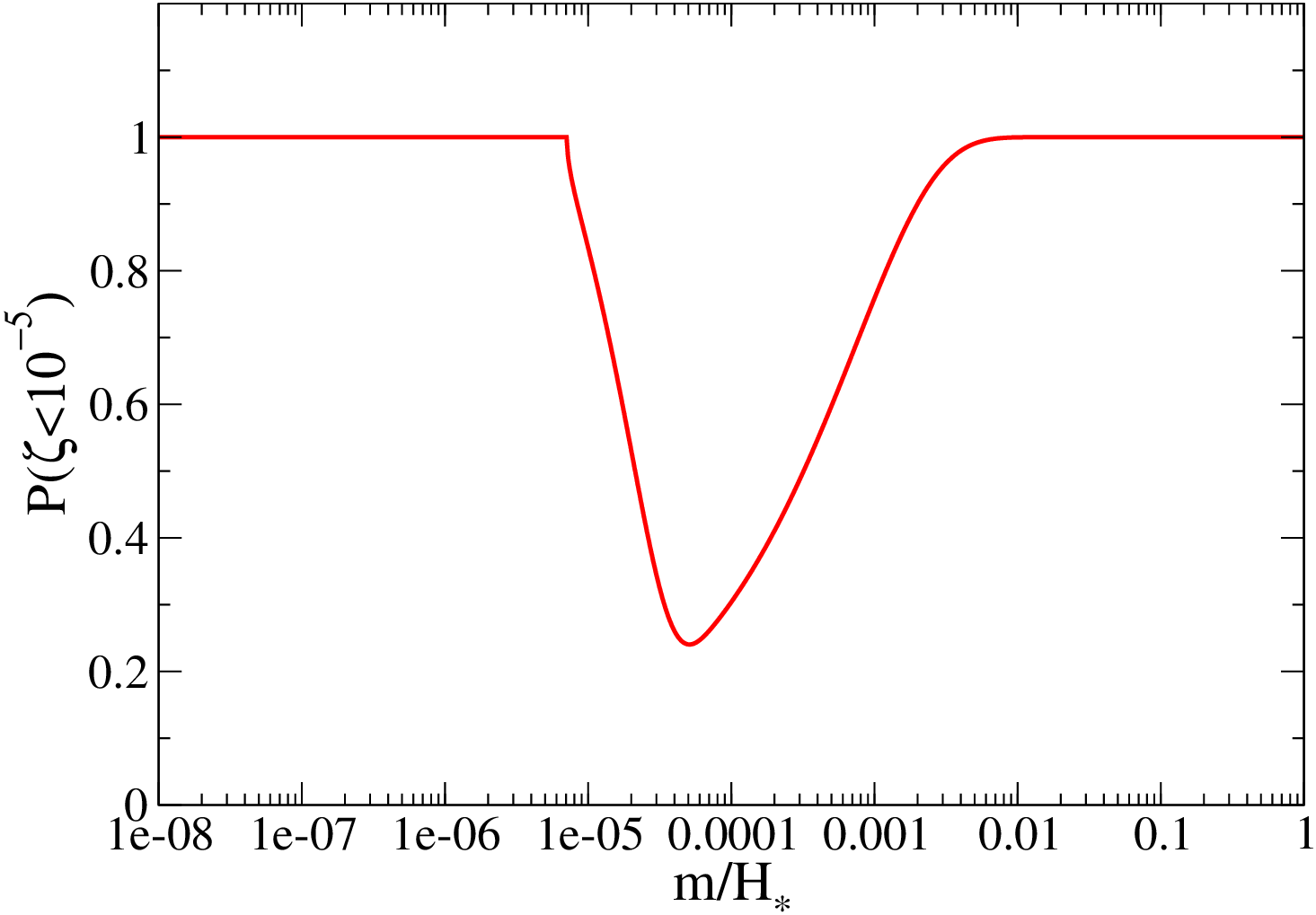}
\caption{The probability of $\zeta<10^{-5}$ as a function of $m$, in the equilibrium distribution. If we require our Universe to have $P(\zeta<10^{-5}) \sim 0.5$, then the figure clearly shows a preferred range for $m/H_*$.}
\label{fig:zetaP}
\end{center}
\end{figure}
%%%%%%%%%%%%%%%%%%%%%%%%%%%%%%%%%%%%%%%%%%%%%%%%%%%%%%%%%%%%%%%%%%%%%%%%
%

Observations indicate that our Universe has $\langle\zeta^2\rangle\simeq (10^{-5})^2$. Any curvaton model with $\zeta \geq 10^{-5}$ is therefore ruled out. Models with $\zeta \ll 10^{-5}$ are not ruled out, but require an additional source of $\zeta$. Figure \ref{fig:zetaP} shows the probability of $\zeta < 10^{-5}$. If our Universe is `typical', we would expect it to reside in the middle of the distribution, so that $P(\zeta<10^{-5}) \sim P(\zeta>10^{-5}) \sim 0.5$. For small $m$, $P(\zeta<10^{-5}) = 1$ because the condition (\ref{eq:critzeta}) is not satisfied for $\zeta \geq 10^{-5}$. In this case, small values of $m$ lead to $r_{dec} \ll 1$ and small $\zeta$. For large $m$, the distribution of $P(\sigma)$ is narrow, and the $\sigma_*^-$ contribution dominates $P(\zeta)$. This favours small $\zeta$ and again gives $P(\zeta<10^{-5}) \approx 1$. However, for intermediate masses, both the $\sigma_*^-$ and the $\sigma_*^+$ contributions are important, and larger values of $\zeta$ are probable. For these intermediate values of $m$, the width of $P(\sigma)$ is such that the maximum of $\zeta(\sigma)$ is favoured. Either larger values of $H_*$ or smaller values of $\Gamma$ cause $\zeta$ to increase, thus changing the range of masses that would give a `typical' Universe.

%%%%%%%%%%%%%%%%%%%%%%%%%%%%%%%%%%%%%%%%%%%%%%%%%%%%%%%%%%%%%%%%%%%%%%%%
\begin{figure}
\begin{center}
\includegraphics[width=0.45\textwidth]{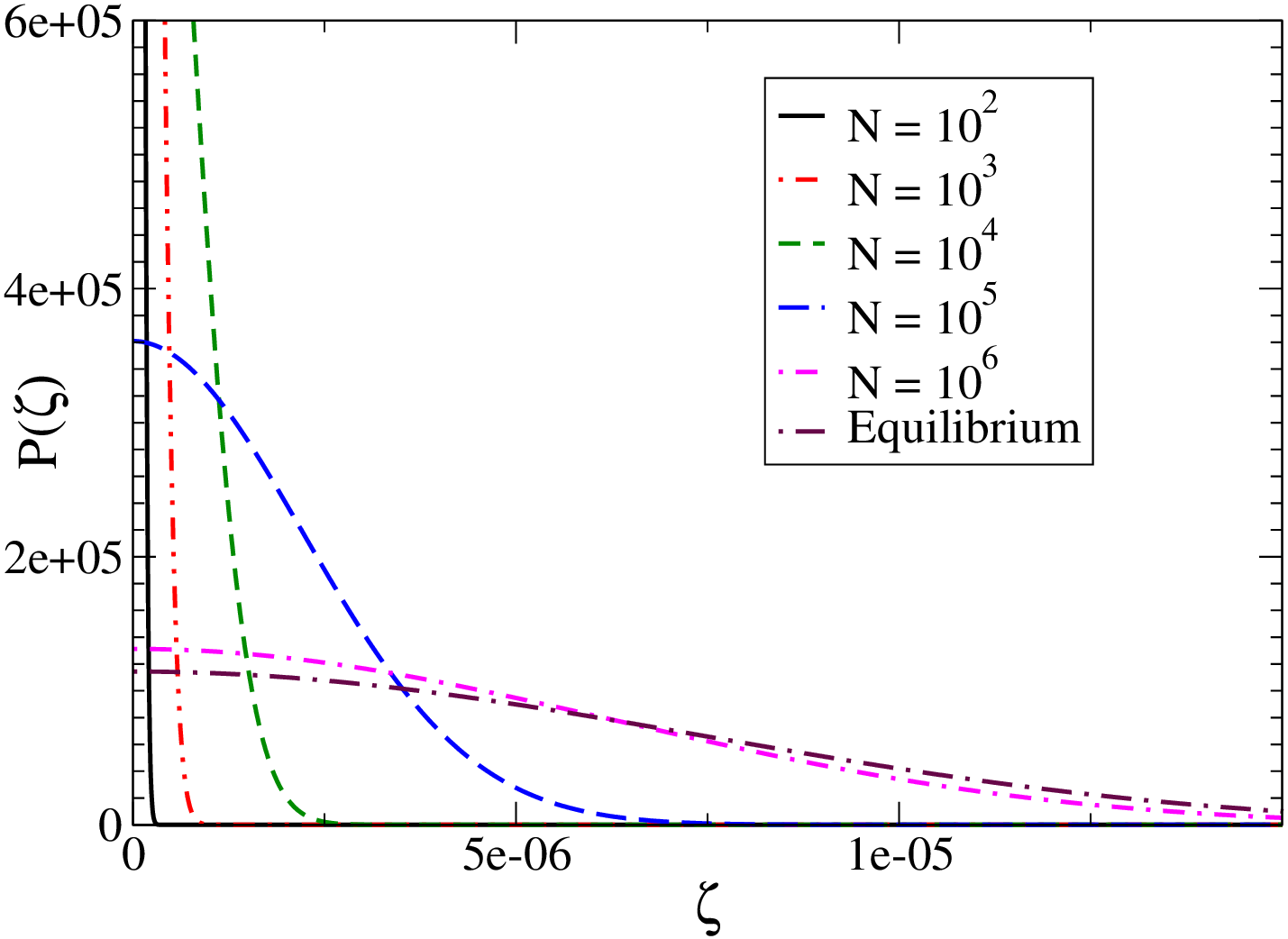}
\includegraphics[width=0.47\textwidth]{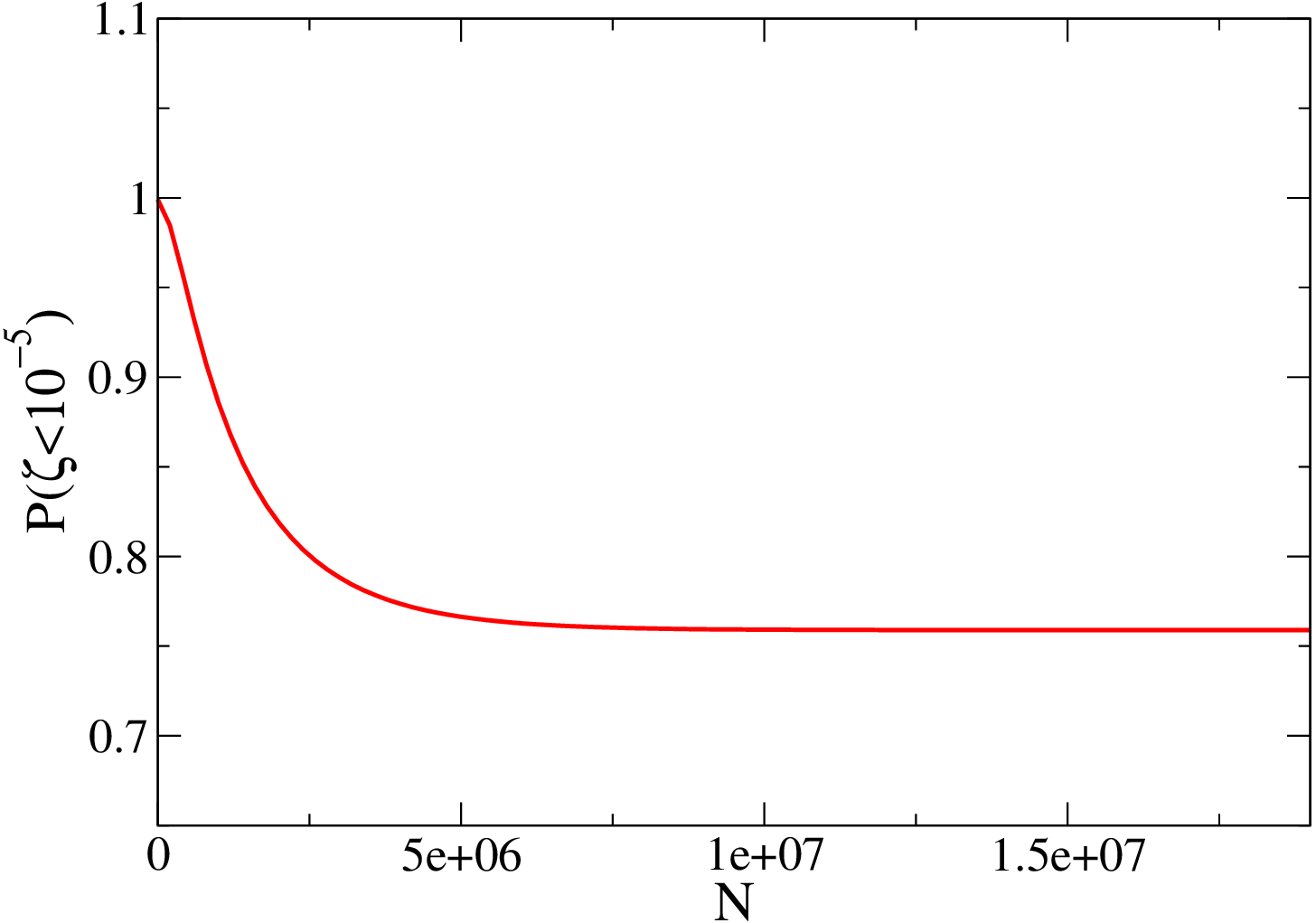}
\caption{Left: The evolution of probability distribution for $\zeta$ for $m/H_*=0.001$ using $H_*=10^{10}$ GeV, $\Gamma=10^{-15}$ GeV. Right: The probability of $\zeta<10^{-5}$ as a function of $N$ for the same parameters as the left panel.}
\label{fig:zetaPN}
\end{center}
\end{figure}
%%%%%%%%%%%%%%%%%%%%%%%%%%%%%%%%%%%%%%%%%%%%%%%%%%%%%%%%%%%%%%%%%%%%%%%%

Let us now consider how the probability distribution of $\zeta$ evolves as a function of the number of $e$-folds. We start with a very narrow Gaussian $w^2(0)\rightarrow 0$ around $\sigma_0=0$, and then let it evolve under the Fokker-Planck dynamics. The initially small width $w^2(N)$ increases towards the asymptotic value, which means that the $\sigma_*$ and thus the $\zeta$ distribution broadens (figure~\ref{fig:zetaPN} (left)). We used $m/H_*=0.001$, so $N_{\rm dec}=1.5\times 10^6$. We see the $\zeta$ distribution widens in time, and moves to the asymptotic in a few million $e$-folds. In figure~\ref{fig:zetaPN} (right) we show the integrated probability for $\zeta<10^{-5}$, where the decoherence timescale is also obvious.

%%%%%%%%%%%%%%%%%%%%%%%%%%%%%%%%%%%%%%%%%
\section{Conclusions}
\label{sec:conc}
%%%%%%%%%%%%%%%%%%%%%%%%%%%%%%%%%%%%%%%%%

We have discussed the evolution of a spectator field in a de Sitter background, going beyond the formal solutions of \cite{Starobinsky}. We have given an explicit solution for the quadratic case, and solved the equation numerically for Gaussian initial conditions for a mixed quadratic and quartic potential. This is important because scalar fields can play a significant role in the evolution of the very early universe. Examples include inflaton fields, moduli fields, quintessence fields, MSSM flat directions and curvaton fields.

Evolution in the case of a quadratic potential is simple as one would expect: if the initial distribution is Gaussian, it will stay Gaussian for the remainder of its evolution. Furthermore, the mean and variance evolve according to exponential laws, and thus all information about the dynamics is encoded in two numbers, $\Nrel$ and $\Ndec$.

However, when one switches on interactions, the dynamics of the system become much more complicated. Even if the distribution starts out Gaussian, it will not stay that way. In fact, all higher order $n$-point functions are non-zero. On sufficiently long timescales the evolution of these correlators is exponential, and we have found the timescales that describe their evolution. Interestingly enough, reaching the exponential behaviour takes about $\Ndec$, before which the distribution evolves in somewhat more complicated ways. This transient behaviour can be surprisingly long, possibly many thousands of $e$-folds. Therefore, unless inflation lasts a very long time, this means that one really needs to investigate the full problem numerically if one wants to study spectator fields.

Furthermore, we should point out that contrary to the general misconception that inflation erases information very quickly, the pre-inflationary conditions of spectator fields are erased very slowly, in many cases taking more than several thousands of $e$-folds.

We have applied these results to the particular case where the spectator field is a curvaton and contributes to the curvature perturbation $\zeta$. We calculated the probability distribution of $\zeta$ for various masses (quadratic potential) in the equilibrium limit. We also showed how the probability distribution of $\zeta$ evolves with time. These arguments could be used to quantify `probable' Universes. It would be interesting to scan the whole parameter space of $\{m, \Gamma, H_*, \sigma_0, N\}$ and discuss the likelihood of $\zeta = 10^{-5}$.

%%%%%%%%%%%%%%%%%%%%%%%%%%%%%%%%%%%%%%%%%
\section*{Acknowledgements}
%%%%%%%%%%%%%%%%%%%%%%%%%%%%%%%%%%%%%%%%%
KE and RL are respectively supported by the Academy of Finland grants 131454 and 218322. OT is supported by the Sofja Kovalevskaja program of the Alexander von Humboldt Foundation and AT is supported by the Carlsberg Foundation.


\begin{thebibliography}{99}

\bibitem{curvorig}
  K.~Enqvist, M.~S.~Sloth,
  %``Adiabatic CMB perturbations in pre - big bang string cosmology,''
  Nucl.\ Phys.\  {\bf B626 } (2002)  395-409.
  [hep-ph/0109214];
  D.~H.~Lyth, D.~Wands,
  %``Generating the curvature perturbation without an inflaton,''
  Phys.\ Lett.\  {\bf B524 } (2002)  5-14.
  [hep-ph/0110002];
  T.~Moroi, T.~Takahashi,
  %``Effects of cosmological moduli fields on cosmic microwave background,''
  Phys.\ Lett.\  {\bf B522 } (2001)  215-221.
  [hep-ph/0110096].

\bibitem{morecurvaton}
For a recent review, see
 A.~Mazumdar and J.~Rocher,
  %``Particle physics models of inflation and curvaton scenarios,''
  Phys.\ Rept.\  {\bf 497}, 85 (2011)
  [arXiv:1001.0993 [hep-ph]].

\bibitem{Caldwell:1997ii}
  R.~R.~Caldwell, R.~Dave and P.~J.~Steinhardt,
  %``Cosmological imprint of an energy component with general equation of state,''
  Phys.\ Rev.\ Lett.\  {\bf 80} (1998) 1582
  [astro-ph/9708069].
  %%CITATION = ASTRO-PH/9708069;%%

\bibitem{EFG}
Kari Enqvist, Daniel G. Figueroa, Gerasimos Rigopoulos,
% Fluctuations along supersymmetric flat directions during Inflation.
JCAP 1201 (2012) 053
[arXiv:1109.3024 [astro-ph.CO]].

\bibitem{Liddle:1998jc}
  A.~R.~Liddle, A.~Mazumdar, F.~E.~Schunck,
  %``Assisted inflation,''
  Phys.\ Rev.\  {\bf D58 } (1998)  061301.
  [astro-ph/9804177].

\bibitem{Dimopoulos:2005ac}
  S.~Dimopoulos, S.~Kachru, J.~McGreevy, J.~G.~Wacker,
  %``N-flation,''
  JCAP {\bf 0808 } (2008)  003.
  [hep-th/0507205].



\bibitem{subir}
  J.~A.~Adams, G.~G.~Ross and S.~Sarkar,
  %``Multiple inflation,''
  Nucl.\ Phys.\ B {\bf 503} (1997) 405
  [hep-ph/9704286].
  %%CITATION = HEP-PH/9704286;%%

\bibitem{Starobinsky86}
  A.~A.~Starobinsky,
  Lect. Notes in Phys., v. 246,
pp. 107-126 (1986).


\bibitem{Nakao:1988yi}
  K.~-i.~Nakao, Y.~Nambu, M.~Sasaki,
  %``Stochastic Dynamics Of New Inflation,''
  Prog.\ Theor.\ Phys.\  {\bf 80 } (1988)  1041.

\bibitem{Stewart91}
  J.~M.~Stewart,
  %``The Stochastic dynamics of chaotic inflation,''
  Class.\ Quant.\ Grav.\  {\bf 8} (1991) 909.
  %%CITATION = CQGRD,8,909;%%

\bibitem{Starobinsky}
  A.~A.~Starobinsky and J.~Yokoyama,
  %``Equilibrium state of a selfinteracting scalar field in the De Sitter background,''
  Phys.\ Rev.\ D {\bf 50}, 6357 (1994)
  [astro-ph/9407016].
  %%CITATION = ASTRO-PH/9407016;%%


\bibitem{selfinteract}
  K.~Enqvist and S.~Nurmi,
  %``Non-gaussianity in curvaton models with nearly quadratic potential,''
  JCAP {\bf 0510}, 013 (2005)
  [astro-ph/0508573];
 K.~Enqvist and T.~Takahashi,
  %``Signatures of Non-Gaussianity in the Curvaton Model,''
  JCAP {\bf 0809}, 012 (2008)
  [arXiv:0807.3069 [astro-ph]];
   K.~Enqvist, S.~Nurmi, G.~Rigopoulos, O.~Taanila and T.~Takahashi,
  %``The Subdominant Curvaton,''
  JCAP {\bf 0911}, 003 (2009)
  [arXiv:0906.3126 [astro-ph.CO]];
  K.~Enqvist, S.~Nurmi, O.~Taanila and T.~Takahashi,
  %``Non-Gaussian Fingerprints of Self-Interacting Curvaton,''
  JCAP {\bf 1004}, 009 (2010)
  [arXiv:0912.4657 [astro-ph.CO]];
  K.~Enqvist, A.~Mazumdar and O.~Taanila,
  %``The TeV-mass curvaton,''
  JCAP {\bf 1009}, 030 (2010)
  [arXiv:1007.0657 [astro-ph.CO]];
 K.~Enqvist, R.~N.~Lerner and O.~Taanila,
  %``Curvaton model completed,''
  JCAP {\bf 1112}, 016 (2011)
  [arXiv:1105.0498 [astro-ph.CO]].
  D.~H.~Lyth,
  %``Non-gaussianity and cosmic uncertainty in curvaton-type models,''
  JCAP {\bf 0606 } (2006)  015.
  [astro-ph/0602285].
  E.~J.~Chun, K.~Dimopoulos, D.~Lyth,
  %``Curvaton and QCD axion in supersymmetric theories,''
  Phys.\ Rev.\  {\bf D70 } (2004)  103510.
  [hep-ph/0402059].






\end{thebibliography}
\end{document}